# EVIDENCE FOR A VERY LARGE-SCALE FRACTAL STRUCTURE IN THE UNIVERSE FROM COBE MEASUREMENTS

Elisabete M. de Gouveia Dal Pino[1], Annibal Hetem[1,2], J.E.Horvath[1],
Carlos A. W. de Souza [3], Thyrso Villela[3] & J.C.N. de Araujo[1,4]

1. Instituto Astronômico e Geofísico, Universidade de São Paulo, Av. Miguel Stéfano, 4200, São Paulo, SP (04301-904), Brazil . E-mail:˝dalpino@plasma.iagusp.usp.br
2. CEA/Saclay, Service d' Astrophysique, 91191˝Gif-Sur-Yvette, France
3. Instituto Nacional de Pesquisa Espaciais, São José dos˝Campos, SP, 12201-970, Brazil
4. S.I.S.S.A., Strada Costiera 11, 34014 Trieste, Italy

Subject headings : cosmology: cosmic microwave background -˝theory - early universe large-scale  structure of the universe

**Abstract**

In this work, we analyse the temperature fluctuations of the cosmic microwave background radiation observed by COBE and show that the distribution can be fitted by a fractal distribution with a fractal dimension D=1.43±0.07. This value is in close agreement with the fractal dimension obtained by Coleman and Pietronero (1992) and Luo and Schramm (1992) from galaxy-galaxy and cluster-cluster correlations up to ~100 $h^{-1}$ Mpc. The fact that the observed temperature fluctuations correspond to scales much larger than 100 $h^{-1}$ Mpc and are signatures of the primordial density fluctuations at the recombination layer suggests that the structure of the matter at the early universe was already fractal and thus non-homogeneous on those scales. This result may have important consequences for the theoretical framework that describes the universe.

## 1. Introduction

Recently, temperature fluctuations of about 13 µK have been detected by the COBE team in the 2.7 K cosmic microwave background radiation (CMBR) over angular scales larger than $7^o$ (Smoot et al. 1992; Bennett et al. 1994). At cosmological distances, an angle $\theta=10^o$, for example, corresponds to a linear size of 1.05 $(\Omega_o h^{-1})$ comoving Gpc (or ~1000 $h^{-1}$ Mpc, assuming a Hubble constant $H_o=100$ h km s$^{-1}$ Mpc$^{-1}$ and $0.5 \leq h \leq 1$) (see e.g., Bertschinger 1994 for a review). These large angular scale fluctuations are believed to be related to primordial gravitational potential fluctuations through the Sachs-Wolfe effect (Sachs and Wolfe 1967). However, the COBE measurements probe only scales which are about more than 7 times larger than the biggest known galaxy superclusters. So, the COBE measurements alone cannot test structure formation theories (e.g., Bertschinger 1994). Measurements on different size scales are being performed to complement the full picture.

It has been generally accepted that on very large scales the distribution of matter is homogeneous and in fact all the existing theoretical approaches are based on this assumption. (e.g. Weinberg 1972). Early analysis of the galaxy distribution which assumed a priori that homogeneity is achieved within the sample size have rendered a galaxy correlation lenght  $r_o \approx 5$ $h^{-1}$ Mpc . According to these analysis galaxies are strongly correlated for $r < r_o$ , while for $r > r_o$ the correlation vanishes rapidly and the distribution becomes homogeneous (e.g., Davis and Peebles 1983). However, the observation of larger scale structures like voids and superclusters on scales up to ~100 $h^{-1}$ Mpc are inconsistent with that "small" value of $r_o$.

Recent re-analysis of galaxy and cluster catalogs by Coleman and Pietronero (1992, hereafter CP) and Luo and Schramm (1992, hereafter LS) show long range (fractal) correlations up to the limits of the samples. The near constant behaviour of the correlation amplitude of the two-point correlation function $\xi(r)$ for clusters indicates that the clustering process may be roughly scale invariant, or in other words, that the structure is fractal. In fact, they claim that the distribution of the visible matter in the universe is fractal or multifractal up to the present observed limits (~100 $h^{-1}$ Mpc) without any evidence for homogenization on those scales. The fractal dimension obtained from the galaxy-galaxy and cluster-cluster correlations is D ~ 1.2 - 1.3″on these scales (CP; LS).

In this work, we perform a statistical analysis of the temperature fluctuations of the CMBR observed by COBE and show that it is also consistent with a fractal structure with fractal dimension D=1.43±0.07. Furthermore, we generate a synthetic space temperature distribution with characteristics similar to those of the observed distribution and derive a filling factor for it in order to get some information about the fractal distribution whose projection we are observing.

In section 2, we give a brief description of the properties of fractal structures, summarize the technique employed in this work and present the results of the fractal analysis of the observed temperature fluctuations of the CMBR. In section 3, we point out our conclusions and the main implications of our results.

## 2. The Fractal Analysis of the CMBR Fluctuations

A review on fractal theory may be found in Mandelbrot (1982) and Falconer (1985) (see also CP and references therein).

A fractal can be described as a system in which more and more structure appears at smaller and smaller scales and the structure at small scales is similar to the one at large scales. This property (called self-similarity) implies the absence of analyticity or regularity everywhere in the system. In general, a fractal can be generated by simple recursive laws which are applied many times on successive″sizes.

Let us consider a two-dimensional fractal. The relation between the perimeter p and the area a of the fractal distribution is (e.g., Hentschel & Proccacia 1984; Hetem & Lépine 1993, hereafter HL):

$$a^{1/2} = F\, p^{1/D} \qquad (1)$$

where D is the fractal dimension and F is the shape factor which is related to the form of the distribution. For fractals, D is in the range $1 < D < 2$.

For our numerical analysis we chose the most sensitive channel (53 GHz) of the COBE/DMR one-year data (Smoot et al. 1992).. The one-year map combines channels A and B at 53 GHz in the same way as described by Smoot et al.

In order to obtain the fractal dimension from the image, we measured the perimeter and the area of regions situated within iso-temperature contours. The total temperature interval of the image was linearly divided in 12 small temperature intervals. The area of each temperature contour region was obtained by counting the pixels above the temperature threshold of that interval, and the perimeter by counting the pixels below the threshold. This method is the same employed by Bazell & Désert (1988) and HL in their analyses of molecular cloud images. The points in a logarithmic plot of perimeter versus area, obtained by varying the threshold temperature along the 12 intervals, are fitted by a straight line in agreement with eq. 1. $D/2$ is the slope, and $F = 10^{-b/D}$, where b is the line intercept. The result is shown in Fig. 1. The corresponding fractal dimension is $D = 1.43 \pm 0.07$ and the shape factor is $F = 0.42 \pm 0.04$. Notice that the errors quoted here are standard errors associated with the least-square fitting procedure.

We also evaluated the fractal dimension of the "noise" of the COBE/DMR map, which is given by the difference between channels A and B signals and found a fractal dimension smaller than unity. This implies that the noise has no self-correlation or similarity property as it would be expected if it were a fractal.

Finally, we have evaluated the putative fractal character of the map by direct calculation of the entropy (as defined by the information theory) through direct counting of the occupation probability of the pixels in non-overlapping circular regions randomly distributed. This has been done for several sets of regions of increasing radius in steps of two degrees. Besides the expected fluctuations we have checked that the mean occupation probability (and therefore the entropy) is the same for any selected size of the region, which in turn means that the entropy does not change when the scale is varied. This is the expected behavior for a true fractal distribution (in fact, we have not found any evidence of multifractality through this entropy analysis) and adds confidence to the validity of the above analysis (a more datailed discussion will be presented˝elsewhere).

Now, we search for fractal models that can reproduce the observed distribution of temperature fluctuations of the CMBR. The technique employed here is based on the comparison of the actual image with simulated images generated by an algorithm which is described in detail by HL (see also Gouveia Dal Pino et al. 1994). In order to generate a synthetic fractal, the following steps are performed: i) a cube with a length edge a=64 and a given maximum temperature T is generated; ii) the cube is divided in eight smaller cubes each one with 1/8 of the original volume; iii) random numbers $s_1$, $s_2$, ..., $s_8$ are attributed to each cube while maintaining the normalization $s_m/s_M = A$, where $s_M$ and $s_m$ are the maximum and minimum values of $s_i$, respectively, and A is the generation parameter; iv) for each small cube, the temperature $t_i$ can be calculated through the relation $t_i \propto T\, s_i$ ; v) for each cube, one successively applies the steps ii) to iv) until the value of the cube edge reaches the minimum value. $0 < A < 1$ is the only parameter of the model and represents the amplitude of variation of the temperatures of the small˝cubes.

A three-dimensional matrix containing the simulated temperature distribution has been constructed. The fractal dimension, as well as the shape factor were obtained from the projection P of the generated matrix V on a plane perpendicular to one of the orthogonal axis that define the matrix V.

A large number of synthetic distributions was constructed by varying the generation parameter A (see Gouveia Dal Pino et al. 1994 for details) and the one with the geometrical characteristics closer to the observed distribution was selected. Fig. 2 shows the contour map of the "best" synthetic distribution we have obtained. The corresponding fractal dimension is D=1.42 and the shape factor is F=0.42, for a generation parameter A=0.1, to be compared with Fig. 1. We notice that, even though we have generated synthetic fractal distributions which are three-dimensional, the method is intrinsically inappropriate for deriving a physical thickness for the last scattering surface because the assumed cubic volume has an arbitrary scale. Fortunately, the effects of a finite thickness on the CMBR maps are not likely to be important for very large scale fluctuations (>> degree scale), although they may be relevant for the smaller scales ( e.g., Dodelson & Jubas 1994).

A filling factor f was evaluated from the generated distribution. It is defined as the ratio between the occupied volume and the total volume of the distribution (notice, however that, due to the self-similarity property of the fractal distribution, the determination of f is actually independent of the volume assumed for the distribution). In order to determine f as a function of the temperatures of the distribution, for a given temperature T, we counted all the volume elements with a temperature above that value. The resulting function is a power-law:

$$f(T) = (0.80 \pm 0.09) \, T^{(1.31 \pm 0.04)} \qquad (2)$$

with a correlation index $d^2 = 0.994$.

## 3. Discussion

The fractal dimension evaluated from the analysis of the observed temperature fluctuations of the CMBR, $D=1.42\pm0.07$, is in reasonable agreement with the fractal dimension evaluated by CP and LS from galaxy-galaxy and cluster-cluster correlations up to ~100 $h^{-1}$ Mpc ($D$ ~1.2-1.3). The temperature fluctuations measured by COBE correspond to length scales larger than ~1000 $h^{-1}$ Mpc. Thus, while the statistical analyses of galaxy and cluster distributions suggest that the distribution of the visible matter is fractal up to ~ 100 $h^{-1}$ Mpc, our results indicate that the temperature (and matter) distributions seem to have been fractals on even larger scales.

The inferred filling factor $f(T)$ (eq. 2), which gives the fraction of the total volume which is filled up by the radiation (and thus by the total matter), provides information about the distribution of the seeds (or primordial density fluctuations) which produced the present large-scale structures of the universe.

Several questions arise as a consequence of the analysis: Does the fractal structure imprinted on the CMBR by the primordial matter density fluctuations at $t_{rec}$, on scales >>100 $h^{-1}$ Mpc, still persist at the present large-scale universe as it seems to occur on smaller scales (as indicated by galaxy and cluster distributions)? If so, how far does the fractal correlation extend or where does the universe become effectively homogeneous? Does the observed fractal structure on large scales at $t_{rec}$ also occurs on smaller scales at that epoch?. We certainly do not have definitive answers to these questions since they will require further extensive work, but we can nevertheless try to draw some tentative conclusions.

We assume, following LS, that some kind of growth process onto primordial seeds provides the fractal correlation while gravity enhances the correlation amplitude on small scales. Numerical N-body simulations exploring the aggregation of matter onto seeds (Frenk et al. 1990) show that, in general, matter undergoes a stochastic motion in space until it is gravitationally bound by seeds to form clumps, and the growth rate of the clump is limited by the diffusing flux of matter onto the seeds (e.g., Witten & Sander 1984). If the aggregate grows by absorbing particles that are randomly moving in a d-dimensional growth space, then Ball & Witten (1984) find that the fractal grown from a diffusion-limited process satisfies $D \geq d - 1$. Thus, our derived fractal dimension $D \approx 1.4$ implies $d < 2.4$ and thus the growth space should involve a 2-dimensional sheet-like object. This conclusion is in agreement with LS analysis. As they pointed out, this fact can constrain the properties of the seeds of the large-scale structure favoring, for example, light domain walls, pancakes, cosmic strings, or superconducting strings seed models (see, e.g., Davis et al. 1992; LS and references therein). There is, however, an important difference between our results and LS discussion: while they argue that the diffusion-limited process of aggregation of matter onto some sort of seeds evolved to a fractal distribution, our results indicate that the seeds themselves were already a fractal at the recombination layer (at least on very large scales). Thus, the growth process should have acted on a fractal distribution of seeds.

Since the growth process is limited by the diffusion of particles onto the aggregate, its rate can become smaller than the expansion rate of the universe at some extension. At this point we should expect the end of the fractal structure in the universe and the beginning of homogeneity. To evaluate the breakdown scale of the fractal correlation (the fractal correlation length L ), LS impose the observed $\delta T/T \sim 10^{-5}$ as the maximum perturbation that can be created by the process. Then, considering a light domain wall as a seed model they use the relation $\delta T/T \; \alpha \; \delta\rho/\rho(H_oL/c)^3$ (Turner et al. 1991) and noting that the fractal growth process occurs only while $\delta\rho/\rho > 1$, they find a lower limit for the fractal correlation length $L \geq 100 \; h^{-1}$ Mpc. This result suggests that the fractal structure obtained in this work for primordial seeds at larger scales ($>1000 \; h^{-1}$ Mpc) has probably been diluted as a consequence of the diffusion-limited growth process in the expanding universe. However, only observations can tell us the actual value of L and thus, where the fractal scale ends up and homogeneity begins. For example, an increase in the maximum observed amplitude of the temperature fluctuations would increase the lower limit of L.

Another interesting aspect is that, as pointed out by CP, the assumption of analyticity (implicit in the Cosmological Principle) is not supported if the matter and radiation happen to be fractals because they introduce an asymmetry between space points. In such a case, a modification of the Cosmological Principle may be necessary (see CP), even more if we have to live with a larger fractal˝correlation length.

Finally, we should emphasize that, while the previous fractal analysis of galaxy and cluster distributions (CP, LS) are dependent on redshift determinations, which are, in turn, obtained under the assumption of a homogeneous universe, the fractal dimension derived in this work is solely determined from the observed CMBR˝fluctuations.

This work was partially supported by the Brazilian Agencies FAPESP and CNPq. We thank J.Nunez and D. Sciama for useful comments and advice. We are also thankful to an anonymous referee for calling our attention to the question of the finite thickness of the last scattering surface.

# REFERENCES

Ball, R.C. and Witten, T.A. 1984, Phys. Rev. A29, 2966

Bazell, D. & Désert, F.X. 1988, Ap.J., 333, 353.

Bennett, C.L. et al. 1994, Ap. J. (in press)

Bertschinger, E.,1994 (preprint)

Coleman, P.H. & Pietronero, L. 1992, Phys. Reps., 213, 311 (CP).

Davis, M. & Peebles, P.J.E. 1983, Ap. J., 267, 465.

Davis, M. et al. 1992, Nature, 356, 489.

Dodelson S. & Jubas, J.M. 1994, Ap. J.

Falconer, K.J. 1985, The Geometry of Fractal Sets (Cambridge Univ. Press, Cambridge).

Frenk, C., White, S., Efstathiou, G., & Davis, M. 1990, Ap. J.,351, 10.

Gouveia Dal Pino, E.M. et al. , in Proceedings of the Vulcano Workshop 1994. Eds. G.Mannocchi & F.Giovanelli, in press.

Hentschel, H.G.E. & Proccacia, I. 1984, Phys. Rev. A., 29, 1461

Hetem, A. & Lépine, J. 1993, A.&A., 270, 451 (HL).

Luo, X. & Schramm, D.N. 1992, Science, 256, 513.

Mandelbrot, B.B. 1982, The Fractal Geometry of Nature (Freeman, S. Francisco).

Sachs, R.K. & Wolfe, A.M. 1967, Ap. J., 147, 73.

Smoot, G.F. et al. 1992, Ap.J., 396, L1.

Turner, M.S., Walkins, R., & Widrow, L. 1991, Ap. J., 367, L43..

Witten, T.A. & Sander, L.M. 1984, Phys. Rev. Lett., 47,1400.

Weinberg, S. 1972, Gravitation and Cosmology (Wiley, N.Y.).

**Figure Captions**

Figure 1. The perimeter vs. area relation obtained from the temperature contours of the observed distribution of the CMBR fluctuations by COBE (see eq. 2).

Figure 2. Temperature contour map of the "best" synthetic distribution of the CMBR fluctuations (contoured from $4.16 \times 10^{-5}$ to $8.42 \times 10^{-4}$ in intervals of $5 \times 10^{-5}$; the isocontour numbers are given in units of $10^{-6}$).